# Cycle-consistent Generative Adversarial Network Synthetic CT for MR-only Adaptive Radiation Therapy on MR-Linac

Gabriel L. Asher[1], Bassem I. Zaki[3,4], Gregory A. Russo[3,4], Gobind S. Gill[4],
Charles R. Thomas[3,4], Temiloluwa O. Prioleau[1], Rongxiao Zhang[2,3,4,5] and Brady Hunt[2,3,4]

[1]Department of Computer Science, Dartmouth College, Hanover, NH, USA
[2]Thayer School of Engineering, Dartmouth College, Hanover, NH, USA
[3]Geisel School of Medicine, Dartmouth College, Hanover, NH, USA
[4]Dartmouth Cancer Center, Dartmouth-Hitchcock Medical Center, Lebanon, NH, USA
[5]Department of Radiation Medicine, Westchester Medical Center, New York Medical College, Valhalla, NY, USA

## Abstract

**Purpose:** To demonstrate the feasibility and accuracy of Deep Learning (DL) based synthetic CT (sCT) generation for dosimetric evaluation in the context of MR-guided adaptive radiation therapy (MRgART).

**Methods:** We trained and evaluated a Cycle-GAN model using paired MRI and deformably registered CT scan slices (dCT) acquired during routine use of an MR-LINAC for MRgART. Synthetic CT volumes were generated from the MR to CT module of the Cycle-GAN network. Data were retrospectively analyzed from treatment plans of patients undergoing treatment for kidney, pancreas, liver, lung, bone, and prostate tumors. Synthetic CT images were assessed using the dCT scans as a reference standard by mean absolute error in Hounsfield Units (HU) and three image similarity metrics: structural similarity index measure (SSIM), peak signal-to-noise ratio (PSNR), and normalized cross correlation (NCC). Synthetic CT volumes were also imported into a clinical treatment planning system and dosimetric calculations re-evaluated for each treatment plan.

**Results:** We trained the model using 8405 frames from 57 patients and evaluated it using a test set of 357 sCT frames from 17 patients. Quantitatively, sCTs were comparable to electron density of dCTs, while improving structural similarity with on-table MRI scans. The MAE between sCT and dCT was 49.2±13.2 HU, sCT NCC outperformed dCT by 0.06, and SSIM and PSNR were 0.97±0.01 and 19.9±1.6 respectively. Furthermore, dosimetric evaluations revealed minimal differences between sCTs and dCTs. Qualitatively, superior reconstruction of air-bubbles in sCT compared to dCT reveal higher alignment between sCT than dCT with the associated MR.

**Conclusions:** Accuracy of deep learning based synthetic CT generation using setup scans on MR-Linacs was adequate for dose calculation/optimization. This can enable MR-only treatment planning workflows on MR-Linacs, thereby increasing the efficiency of simulation and adaptive planning for MRgART.

**Nomenclature**:

DL: Deep learning

MR-LINAC: MRI-guided linear accelerator

dCT: Deformed CT

sCT: Synthetic CT

# Introduction

MR-guided adaptive radiation therapy (MRgART) is a relatively new approach to radiation therapy (RT) which combines the advantages of magnetic resonance imaging (MRI) including improved soft tissue contrast, non-ionizing radiation and real-time imaging with the precise targeting of radiation therapy. By using MRI linear accelerator systems (MRI-linac) during treatment, MRgART allows for real-time tracking of the target, decreased planning margins and prescription dose escalation, which, in combination, can lead to improved outcomes and reduced side effects for patients.[1] Several studies have demonstrated the benefits of MRgART, including improved target coverage, reduced toxicity, and improved overall survival.[2–4] Additionally, MRgART has been shown to be effective in treating a variety of cancer types, including brain, prostate, and breast cancer.[1,5–7] Overall, MRgART is an innovative approach to radiation therapy that has the potential to improve the patient outcomes.

However, a critical weakness of MRgART is a reliance on electron density maps which are derived from computed tomography (CT) images for dose planning. Thus, to successfully carry out MRgART, MR images must be co-registered with CT images. The co-registration of CT and MRI images is a critical step in MRgART which aligns the CT and MRI images allowing for precise targeting of the tumor. However, due to the characteristic differences in the imaging modalities, co-registration inherently introduces uncertainty in geospatial awareness. Moreover, co-registration is conventionally performed manually, which can also be a source of errors if not done carefully. Complications in co-registration can ultimately impact treatment efficiency and radiation plan dosimetry. For example, this co-registration process has a systematic uncertainty of approximately 2-5mm[9]. The errors that can occur during the co-registration process can persist at multiple levels of the treatment workflow and bring systematic errors.[10] Structural discrepancies between the CT and MR scans tends to increase with interval between the sequential acquisitions, which lead to the usual lengthy workflow, anatomical uncertainties and manual corrections such as the tissue/air override which are labor-intensive. Furthermore, CT images have limited soft tissue contrast, whereas MRI images have a better soft tissue contrast profile- provide dynamic and physiological information about the tumor. However, they are sensitive to motion artifacts and the presence of metallic objects and contrast agents[1,8].

Current MR to CT registration techniques involve generating a deformed CT (dCT) from a baseline MR image and an existing CT image of the patient. Many of the conventional CT image registration techniques and following evaluations are time consuming.[11] Computational time for atlas-based methods rises linearly with dataset size and bulk segmentation requires longer acquisition time compared to conventional MR sequences[12,13]. Thus, deep learning (DL) based synthetic CT has been touted as a promising alternative to previous registration techniques with advantages of increased efficiency and eliminated anatomical discrepancies.

However, despite these advantages, there are several significant hurdles for training generalizable deep-learning synthetic CT generation models. First and foremost, datasets are often small. Training machine-learning (ML) models requires ground truth labels, which means that for a MR to CT registration task, paired CT/MR datasets are needed. Thus, datasets used in training synthetic CT models tend to be small by machine learning standards and can be afflicted by batch errors due to limited diversity in acquisition settings. Additionally, many datasets suffer from data leakage due to slice-level or fraction-level dataset splits, which inflates model performance. [16,17] Finally, large differences between intensity values from different MRI manufacturers means that trained models struggle to generalize.

Many studies exist evaluating performance of various deep-learning architectures on synthetic CT registration from MR images. Generative adversarial networks (GAN) are among the more popular techniques for mapping MR images to sCTs [18]. A GAN is a type of machine learning model consisting of two neural networks, a generator and a discriminator, that compete against each other to generate realistic and high-quality synthetic data. However, GANs require strongly paired ground truths to train properly. This poses an issue in the context of dataset generation since MR images and CT images cannot be captured simultaneously. Thus, CycleGANs have shown themselves to be a promising method for synthetic CT generation. Due to a cycle-consistency loss, CycleGANs can train on paired or unpaired data, greatly increasing dataset size. Results from related CycleGAN studies have shown promise, with mean absolute error between ground-truth CT volume and sCT volumes between 30-150 Hounsfield Units (HU). [19–23] However, these studies demand further investigation for the following reasons. Firstly, most prior models were only trained with one or two sites in mind. Lei et al. study the brain and pelvis[22], Farjam et al. the pancreas[24], Wolterink et al the brain[21], and Yang et al study the brain and abdomen (with different models for each)[19]. Furthermore, due to the limited size of datasets, many prior studies, such as Kang et al, Farjam et al, and Lei et al, do not use a held out validation data set in addition to their final test set.[22–24] The lack of this validation set means that these models implicitly overfit on the test data since hyperparameters can be directly tuned on test data. Finally, although some studies evaluated treatment planning dose volumetrics on synthetic CTs[23], many of these studies did not[19–22,24].

Thus, in this research study, we aim to further evaluate the performance of CycleGAN, a deep-learning algorithm, on synthetic CT generation. We hypothesize that CycleGAN for synthetic CT generation is at least as effective for MRgART as deformable registration both qualitatively and quantitatively. To address this hypothesis, we employ a novel, large, paired image dataset with 6 different sites. Furthermore, we employ a rigorous data splitting regime to ensure minimal data leakage and the most generalizable results possible. Finally, we use the model trained from this dataset to investigate two tasks. Firstly, we assess the performance of CycleGAN in generating synthetic CT images that accurately correspond to their ground truths. Secondly, we evaluate our synthetic CT images with treatment planning software to determine whether they have adequate dosimetric outcomes to enable MR-only planning on MR-linacs.

## Methods and materials

*Study Dataset*

This study analyzed patients undergoing stereotactic body radiation therapy using the ViewRay MRIdian MR-LINAC at the Dartmouth-Hitchcock Medical Center (DHMC). Patients underwent radiation therapy between March 2021 and June 2022. All DICOM and treatment delivery records were retrospectively accessed and anonymized before inclusion in the study in accordance with a protocol approved by the DHMC institutional review board. Simulation CT/MR scans were acquired acquired on the same day at the outset of RT treatment planning, typically one hour between scans with CT scanning first. CT scans were acquired using a simulation scanner (Siemens EDGE) using routine clinical settings. MRI scans were obtained using a MR-Linac (ViewRay MRIdian) using built-in clinical protocols. Ground-truth deformed CTs (dCT) were generated using the ViewRay treatment planning system registration pipeline.

*Image Pre-Processing and Data Partitioning*

After dCT generation, all dCT and MR images were extracted in DICOM format with 144 slices and a 3mm axial resolution. The in-plane dimensions were 310 x 360 pixels with a 1.5mm resolution. DICOM volumes were converted to tiff images using the python open-source pydicom and tifffile packages. After conversion to tiff format, all images were padded to 440x440 pixels. Next, we conducted a stratified k-fold data splitting scheme using treatment site specified in the DICOM data as the category for stratification including: adrenal, pancreas, liver, lung, bone, prostate, and other. This split yielded 58 training patients, 11 validation patients, and 17 testing patients, ensuring that train and test splits contained at least one patient from each treatment site category. Prior to use in model training, all 8405 CT and MR images were normalized. CT images were normalized with the following linear formula:

$$Y = \frac{X - X_0}{\alpha} \qquad \text{(Eq. 1)}$$

where $X$ is the input voxel in HU, $X_0$ = -1024, and $\alpha$ = the 99.99 percentile of HU values in the training set range. To ensure that all normalized values were positive, $X_0$ was set to -1024, considering that the minimum HU value in our training set was also -1024. MR images were also normalized with Eq.1, with

$X_0$ set at 0, since MR intensity values are all already nonnegative. Our pre-processing pipeline is described in **Figure 1**.

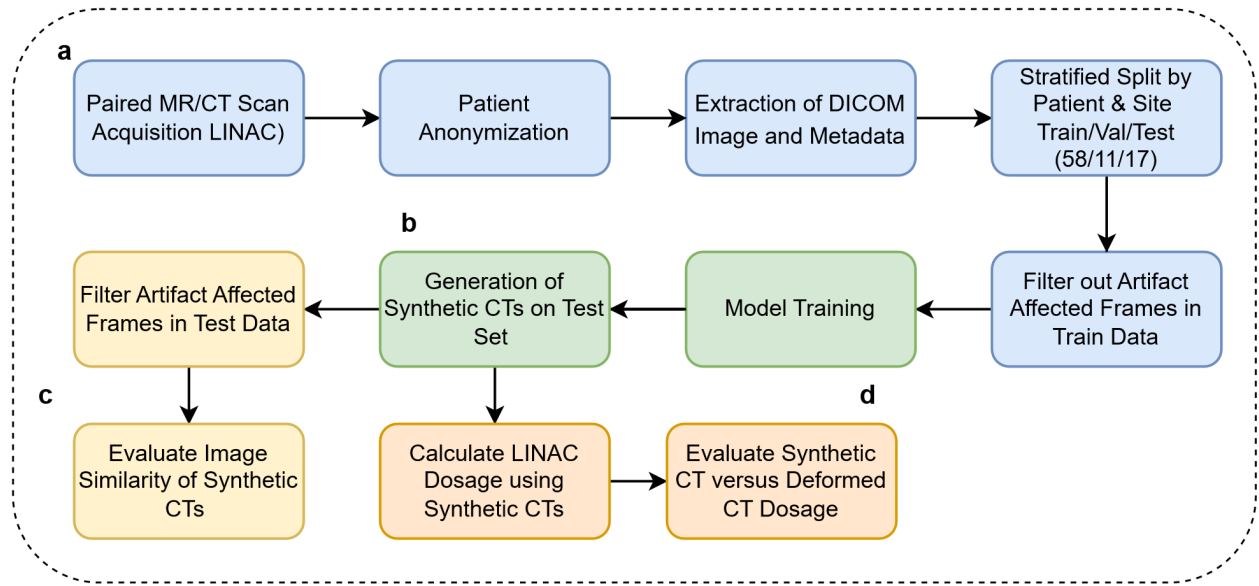

**Figure 1:** a) DICOM anonymization, preprocessing, and filtration b) Model training and inference calls c) Evaluation of model performance on Hounsfield Unit level d) Clinical evaluation of synthetic CTs.

## Model and loss formulation

Our study utilized the CycleGAN (Cycle-Consistent Generative Adversarial Network) model architecture[25]. **Figure 2** details the architecture of CycleGAN. This architecture is a derivative of the generative adversarial network (GAN)[26], a popular deep-learning architecture which leverages two competing networks: a generator and a discriminator. In GANs, a generator module creates a synthetic image from a real input, and the discriminator predicts whether this synthetic image is real of fake. The model stops learning when the generator produces indistinguishable images from the ground truth.

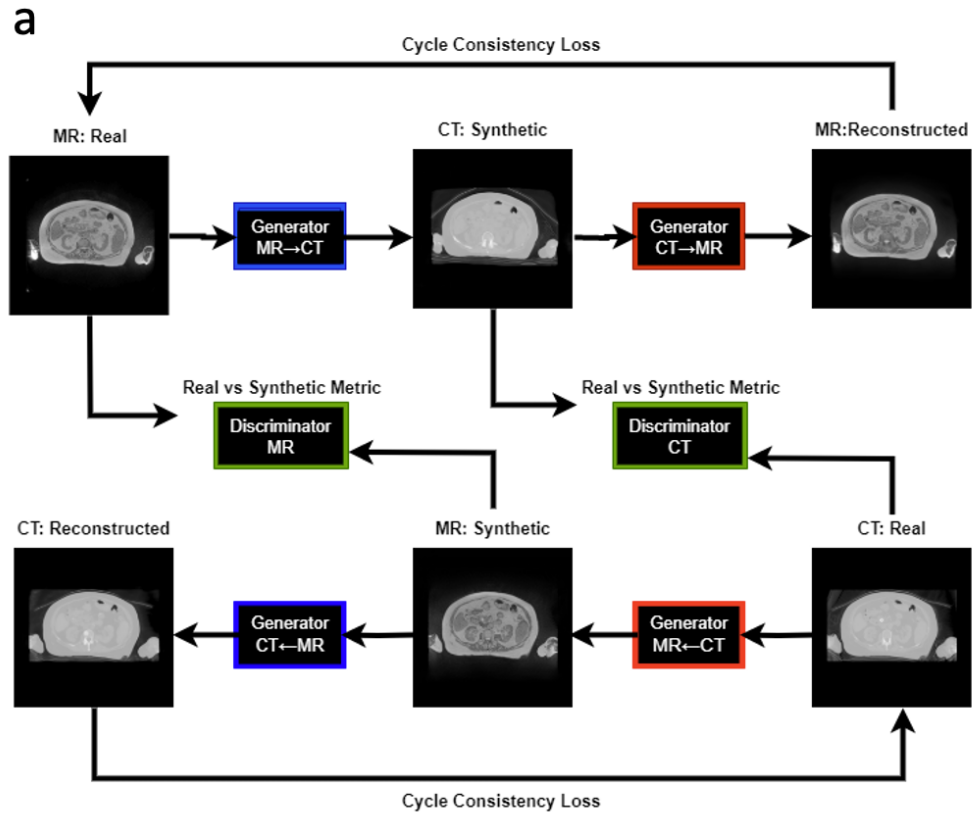

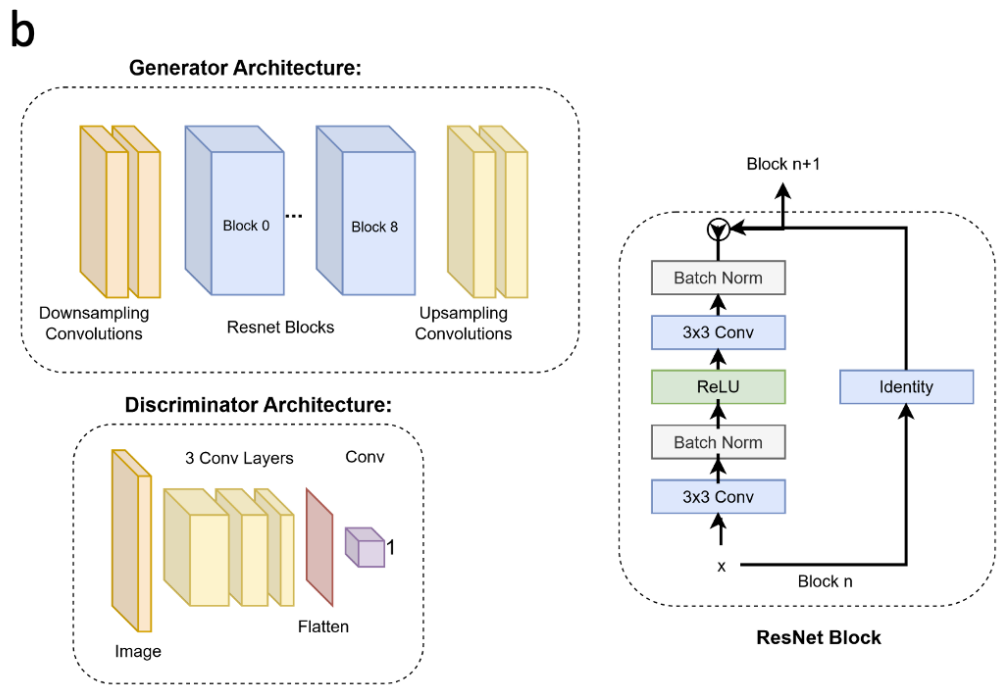

**Figure 2:** Deep learning architecture. a) Flow diagram of paired generator and discriminator modules illustrating MR/CT conversions. b) Block diagrams of layers utilized in generator and discriminator modules.

CycleGAN adopts a similar architecture as a GAN with two key differences. Firstly, instead of having one generator and one discriminator, CycleGAN has two generators ($G_{MR}, G_{CT}$) and two discriminators ($D_{MR}, D_{CT}$). The generator $G_{CT}$ takes an MR image and generates a CT image, while the generator $G_{MR}$ takes a CT image and generates an MR image. The loss functions used to train the generators and discriminators in a Cycle-GAN typically include three components: the adversarial loss, the cycle-consistency loss, and the identity loss. Adversarial loss is used to ensure that the generated images are realistic and can fool the discriminators. The adversarial loss, $L_{adv}$ is calculated using the mean squared error between the discriminator's output and the ground-truth label. Cycle-consistency loss, $L_{cyc}$, is used to ensure that the generated images preserve the content of the original images and is calculated as the mean absolute error (MAE) between the original image and the reconstruction of the original image in the original domain. The cycle loss, $L_{cyc}$ for the generator $G_{CT}$ is calculated as:

$$L_{cyc} = \|G_{MR}(G_{CT}(CT_{gt})) - CT_{gt}\| \cdot \lambda \qquad (Eq.\ 3)$$

where $\lambda$ is a weight, and $CT_{gt}$ is a ground truth CT. The identity loss is added to ensure that the images from the same domain should not change after passing through the generator. The identity loss, $L_{idt}$ for $G_{CT}$ is:

$$L_{idt} = \|G_{CT}(CT_{gt}) - CT_{gt}\| \qquad (Eq.\ 4)$$

The loss formulations for $G_{MR}$ are formulated similarly as $G_{CT}$. Finally, the total loss, $L_{total}$ is calculated as follows:

$$L_{total} = L_{advMR} + L_{advCT} + L_{cycMR} + L_{cycCT} + L_{idtMR} + L_{idtCT} \qquad (Eq.\ 5)$$

We trained our model on 1 Nvidia RTX 2080 Ti with 12GB of GPU memory. Additionally, we used the following model parameters: random crop to 256x256, batch size of 1, 100 epochs. Our model training code was from: https://github.com/junyanz/pytorch-CycleGAN-and-pix2pix [25,27].

*Evaluations*

After model training, we used the lowest loss model (measured on the validation set) to generate a set of 2464 synthetic CT images from a held-out test set. However, whilst evaluating image similarity, we found a large amount of our test set dCTs were corrupted by artifacts in the beginning and end of each series in the axial plane. Thus, we only conducted our evaluations on the dosimetric relevant images within a 3cm radius of the PTV.

*Synthetic Image Quality Assessments*

In this study, we opted for a multifaceted evaluation of image quality using multiple metrics - MAE, PSNR, SSIM, and NCC - each offering unique insights into different aspects of image translation quality. The Mean Absolute Error (MAE) allows us to quantify the average prediction errors, providing a straightforward understanding of the overall translation error. Peak Signal-to-Noise Ratio (PSNR) helps to measure the maximum potential fidelity of the translated image in relation to corruption noise, which is valuable in assessing the strength and clarity of the signal being studied. The Structural Similarity Index Measure (SSIM) is specifically adopted for its sensitivity to changes in structural information. Lastly, Normalized Cross-Correlation (NCC) is used to assess the pattern similarity between the translated and target images. This comprehensive approach ensures a thorough evaluation, as relying on a single metric could yield an incomplete picture of the performance of our deep learning-based MR to CT translation method. The formulas for these metrics are as follows:

$$MAE = \frac{1}{n}\sum_{i=1}^{n} |y_{defCT} - y_{synCT}| \qquad \text{(Eq. 6)}$$

$$PSNR = 10 \cdot \frac{max(y_{defCT})^2}{\frac{1}{n}(\sum_{i=1}^{n}(y_{defCT} - y_{synCT})^2)} \qquad \text{(Eq. 7)}$$

$$SSIM(x,y) = \frac{(2\mu_{ix}\mu_{iy} + C_1)(2\sigma_{ixiy} + C_2)}{(\mu_{ix}^2 + \mu_{iy}^2 + C_1)(\sigma_{ix}^2 + \sigma_{iy}^2 + C_2)} \qquad \text{(Eq. 8)}$$

For SSIM, $\mu_{ix}$, $u_{iy}$ are the average HU values for the ith axial slice of our dCT and sCT series respectively. $\sigma_{ix}^2$, $\sigma_{iy}^2$ represent the variance of the aforementioned dCT and sCT slices. Finally, $C_1$, $C_2$ are constants applied as suggested by Wang et al[28]. Additionally, we evaluated the similarity of sCT and dCT images to their corresponding MR inputs using Normalized Cross-Correlation (NCC). Normalized cross correlation (NCC) is a similarity measure that ranges from -1 to 1, used to determine the degree of similarity between two image regions, with 1 being most similar. NCC was calculated using the xcdskd package on python.

*Dosimetric Assessments*

The second evaluation criteria of our sCTs consists of a comparison between the RT dose calculation in the sCT versus the dCT. To calculate these differences, our tiff images were reconverted to DICOM format. This was done by replacing the dCT DICOM "Pixel Data" tag with our generated sCT image.

Next, we fed our DICOM sCTs into the ViewRay treatment planning software (version 2.0.6) with the same parameters used on the dCTs to create dose volume histograms (DVHs). These DVHs were used to calculate the absolute difference of dose delivered to the PTV at 95%, 90%, and 85% of the volume. Additionally, we also calculate the difference in dose delivered above 33Gy to all 3cm OAR sites.

## Results

### Image quantitative comparisons

**Table 1** reports the mean absolute error (MAE) of our cycle-GAN model. In total, 357 synthetic CT frames were analyzed from 17 patients. Of the treatment sites analyzed, the most common sites were liver, pancreas, and lung, with 84, 105, and 63 frames respectively. MAE was calculated by comparing synthetic CTs versus deformed CTs. Median MAE values across sites ranged between 44.7.4 HU to 55.6 HU with an overall median MAE of 49.2 HU. Pancreas and lung scans had the highest MAE standard deviations with 20.1 and 9.8 HU respectively. Kidney showed the least difference in MAE between scans with a standard deviation of 1.2 HU.

| Value | OTHER | KIDNEY | PANCREAS | LIVER | LUNG | BONE | PROSTATE | ALL |
|---|---|---|---|---|---|---|---|---|
| Median MAE (STD) | 49.4 (4.3) | 44.6 (1.2) | 47.6 (20.0) | 44.3 (7.7) | 55.6 (9.8) | 55.2 (1.6) | 51.4 (1.8) | 49.2 (13.2) |
| Number of Frames | 42 | 21 | 105 | 84 | 63 | 21 | 21 | 357 |
| Number of Patients | 2 | 1 | 5 | 4 | 3 | 1 | 1 | 17 |

**Table 1:** MAE values in Houndsfield Units (HU) of sCT compared to dCT stratified by treatment site.

**Table 2** reports SSIM and PSNR metrics, as well as the NCC of our sCTs and dCTs versus our MR image ground truths. Our sCTs demonstrate a higher (better) NCC value in comparison to deformed CTs in all but one treatment site (Other). Although differences were generally minimal, sCT NCC scores on bone, prostate, and liver showed the largest improvements compared to our deformable registration ground truths. Additionally, we report high SSIM values across all sites, with a median SSIM value of 0.971. Thus, the structural similarity between our sCTs and dCTs is near perfect. Additionally, we report a median PSNR of 19.9 across all sites. This high PSNR value indicates that the sCT is a good representation of the dCT.

| Value | OTHER | KIDNEY | PANCREAS | LIVER | LUNG | BONE | PROSTATE | ALL |
|---|---|---|---|---|---|---|---|---|
| dCT NCC | **0.769** | 0.825 | 0.827 | 0.839 | 0.826 | 0.803 | 0.707 | 0.818 |
| sCT NCC | 0.763 | **0.826** | **0.830** | **0.847** | **0.830** | **0.812** | **0.718** | **0.824** |
| Median SSIM (STD) | 0.974 (0.002) | 0.978 (0.002) | 0.969 (0.016) | 0.974 (0.008) | 0.967 (0.015) | 0.965 (0.003) | 0.976 (0.002) | 0.971 (0.013) |
| PSNR (STD) | 20.3 (0.98) | 20.8 (0.76) | 19.6 (2.05) | 19.9 (1.54) | 19.3 (1.51) | 19.0 (0.17) | 20.4 (0.33) | 19.9 (1.64) |

**Table 2:** Image quality metrics per site. NCC: normalized-cross-correlation, SSIM: structural similarity index measure, PSNR: peak signal to noise ratio.

**Figure 3** presents selected images from the scans with the lowest, media, and highest MAE (HU) from our test set. As is evident in all three examples, our model struggled to properly predict skin, bone, and limbs outside of the torso region (arms). Difficulty predicting bone is consistent with prior works[21,22]. Additionally, we view differences between the sCT and dCT with regards to air bubbles. For example, in the median HU images, we observe that our synthetic CT images correctly include air bubbles present in MR, whereas the deformed CTs do not include said bubbles. However, from a qualitative perspective, our sCT reconstructions are minimally different from their dCT counterparts.

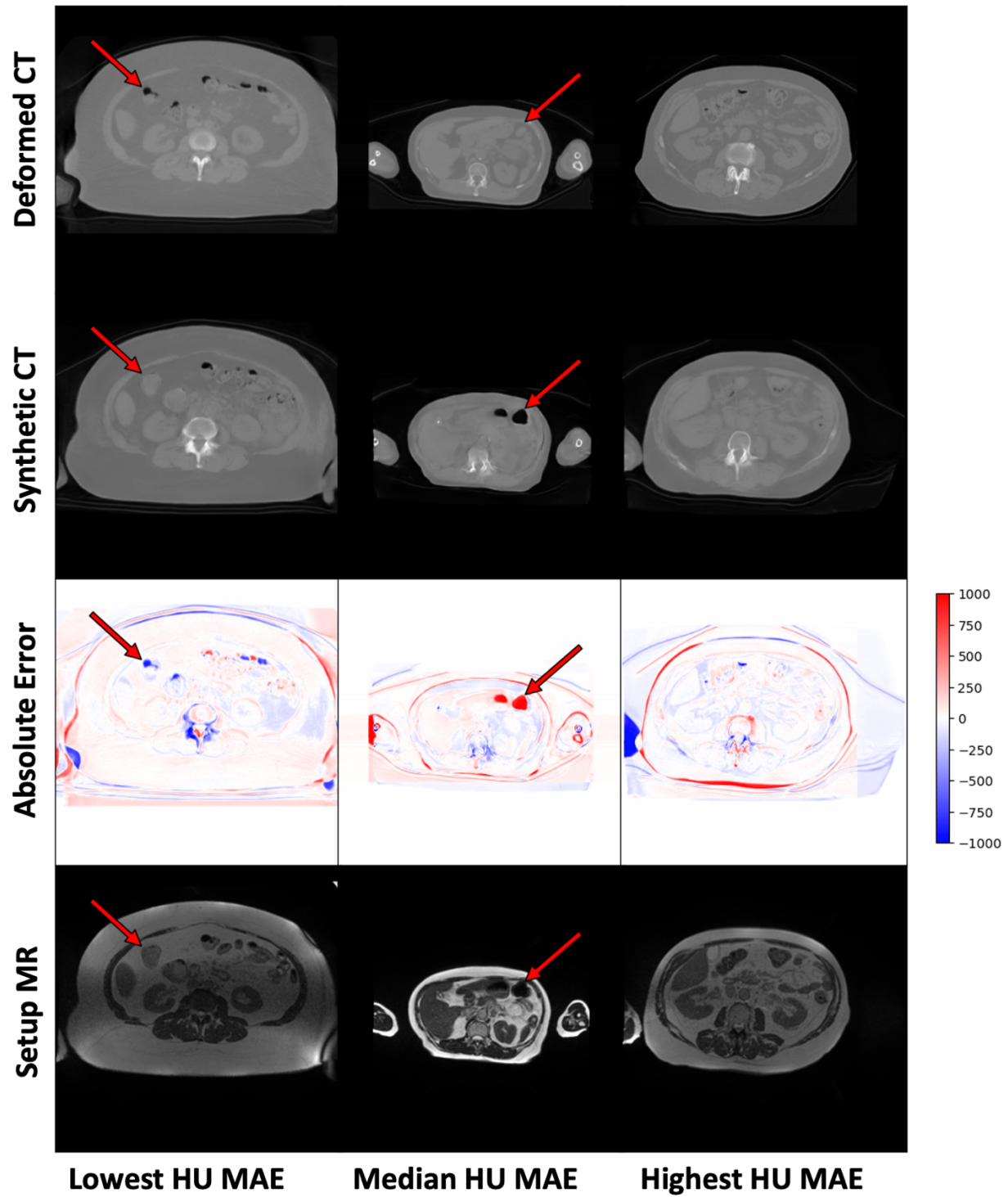

**Figure 3:** The lowest, median, and highest synCT mean average errors (HU) in our test set. Colorbar units are in Houndsfield Units, and red arrows point towards differences in air-bubble reconstruction between sCT and dCT.

*Dose comparisons*

**Figure 4** reports the absolute difference of RT dose delivered to PTV at 85%, 90%, and 95% of the volume based on RT dose recalculations using our sCTs. Our results indicate minimal differences in dose delivered using sCTs. Median difference in dose delivered to PTV across all sites is .45Gy, .47Gy, and .46Gy for 85%, 90%, and 95% of volume respectively. Additionally, we found that dose above 33Gy delivered to OARs within 3cm of the PTV also showed minimum differences between sCT and dCT RT dose calculations. The median difference in dose delivered to 3cm OARs minimally increased by 0.01 Gy after switching from sCT to dCT. Additionally, 4 patients had no change in dose delivered to these sites, and 6 patients had less dose above 33Gy delivered to these sites.

a)

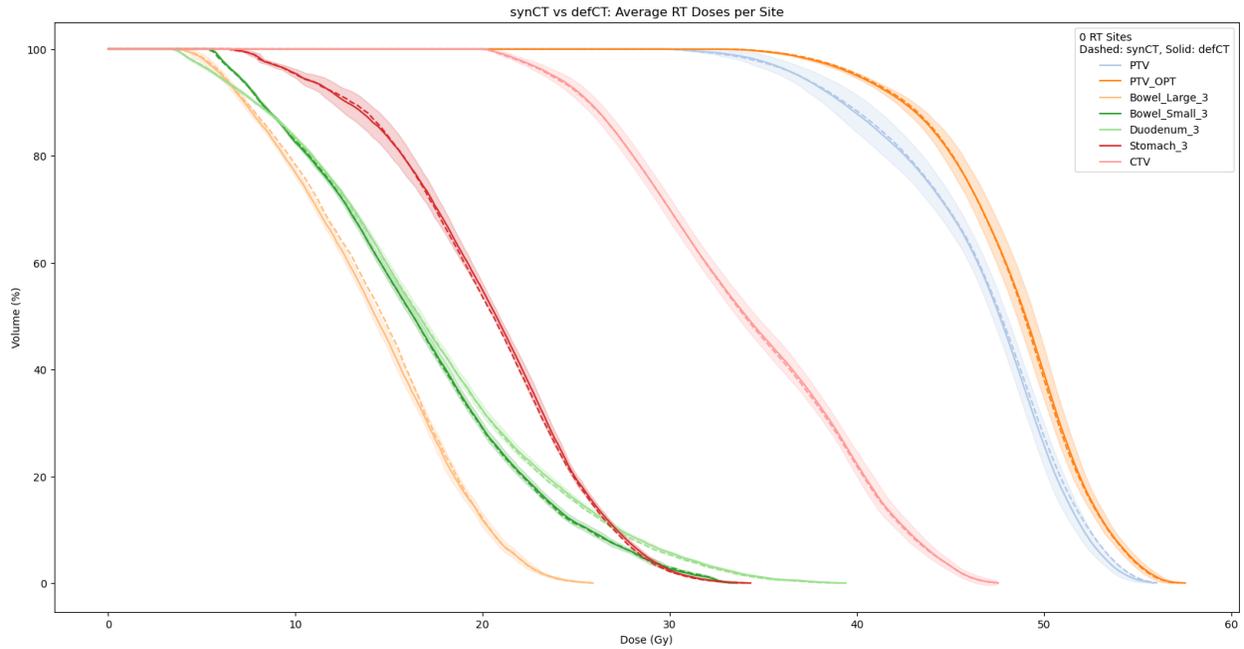

b)

| Patient | Site | Gy Diff @ 85% | Gy Diff @ 90% | Gy Diff @95% | Δ Dose > 33Gy |
|---------|------|---------------|---------------|--------------|---------------|
| PT0049 | OTHER | 0.19 | 0.21 | 0.23 | 0.00 |
| PT0059 | KIDNEY | 0.50 | 0.49 | 0.47 | -0.16 |
| PT0066 | PANCREAS | 0.23 | 0.14 | 0.09 | 0.07 |
| PT0067 | LIVER | 0.00 | 0.06 | 0.01 | 0.00 |
| PT0069 | LIVER | 0.48 | 0.56 | 0.47 | 0.00 |
| PT0070 | PANCREAS | 1.56 | 1.16 | 1.21 | -0.26 |
| PT0074 | LIVER | 0.04 | 0.04 | 0.05 | 0.00 |
| PT0075 | LIVER | 0.86 | 0.82 | 0.28 | -0.18 |
| PT0075 | PANCREAS | 0.70 | 0.70 | 0.45 | 0.65 |
| PT0077 | LUNG | 0.16 | 0.04 | 0.42 | 0.98 |
| PT0078 | PANCREAS | 0.12 | 0.07 | 0.01 | -0.19 |
| PT0080 | PANCREAS | 0.19 | 0.12 | 0.21 | 0.08 |
| PT0081 | BONE | 0.57 | 0.57 | 0.41 | Na |
| PT0082 | PROSTATE | 0.16 | 0.16 | 0.17 | 0.25 |
| PT0083 | LUNG | 1.23 | 1.65 | 1.76 | 0.11 |
| PT0084 | LUNG | 0.25 | 0.68 | 1.13 | -1.11 |
| ALL | ALL | 0.45 | 0.47 | 0.46 | 0.01 |

**Figure 4:** Dosimetric evaluations using sCT. a) Averaged sCT vs dCT dose volumetrics per site (dotted = sCT, solid = dCT). The shaded region corresponds to the standard deviation between the sCT and dCT

DVHs. b) DVH metrics tabulated by patient. Absolute difference in dose delivered between sCT and dCT to 85%, 90%, and 95% of treatment volume. Final column reports the difference in dose delivered above 33Gy in 3cm OAR sites, where a negative value indicates that less dose above 33Gy was delivered with the sCT.

## Discussion

This report presents strong results supporting an MRaGRT workflow. From a synthetic image quality perspective, our CycleGAN implementation reports comparable or superior MAE compared to prior studies conducted using a CycleGAN architecture [19,29]. Additionally, our results indicate that CycleGAN architecture generalizes well to several treatment sites with minimal additional training data. For example, despite only having 1 kidney, bone, and 3 prostate series each in our training data, we still report a mean MAE of 50.4 HU across these sites. Further, our overall median MAE of 49.2 HU is a strong result in comparison to prior work. This improved synthetic image quality may be a result of a larger dataset. Of the prior CycleGAN studies performed, this study analyzed 86 patient scans, compared with 24, 45, and 38, patients in works from Wolterink et al [21], Yang et al[19], and Brou et al[29]. Thus, given our superior reconstruction results, we demonstrate that dataset size is paramount in creating strong generative synthetic CT models.

Our treatment dose comparison also supports an MRgART workflow. Our .45Gy, .47Gy, and .46Gy average differences in dose delivered to the PTV for 85%, 90%, and 95% of the volume indicates that sCTs have minimal effects on dose delivered. Additionally, NCC comparisons in table 2 indicate that synthetic CTs capture an equivalent to better representation of the MR scans taken during treatment. Thus, we can extrapolate that synthetic CT scans may present a more precise image to calculate RT dose with.

Qualitative analysis of images also supports a MRgART workflow. For example, analysis of the air bubbles in **Figure 3** suggests that the synthetic CT better models air pockets within patients, a key limitation of deformed CT methods. Furthermore, visual examination of the sCT and dCT images reveals only minimal differences between the two images. In fact, the areas with the largest differences in HU values in the images are often periphery areas such as arms. This may inflate our MAE value since these periphery areas are part of MAE calculations. Given these observations, we believe that the synthetic CT is likely a better representation to use when performing dose calculations.

A key strength of using a deep learning model in synthetic CT generation is speed. Conventional image registration techniques rely on an iterative image update process. However, this process is slow and requires lengthy computation, which is a bottleneck when performing real-time dose calculation. On the other hand, our deep-learning model is fixed after training and requires a single forward pass to generate an output. Therefore, our DL model is much more suited to real-time RT strategies.

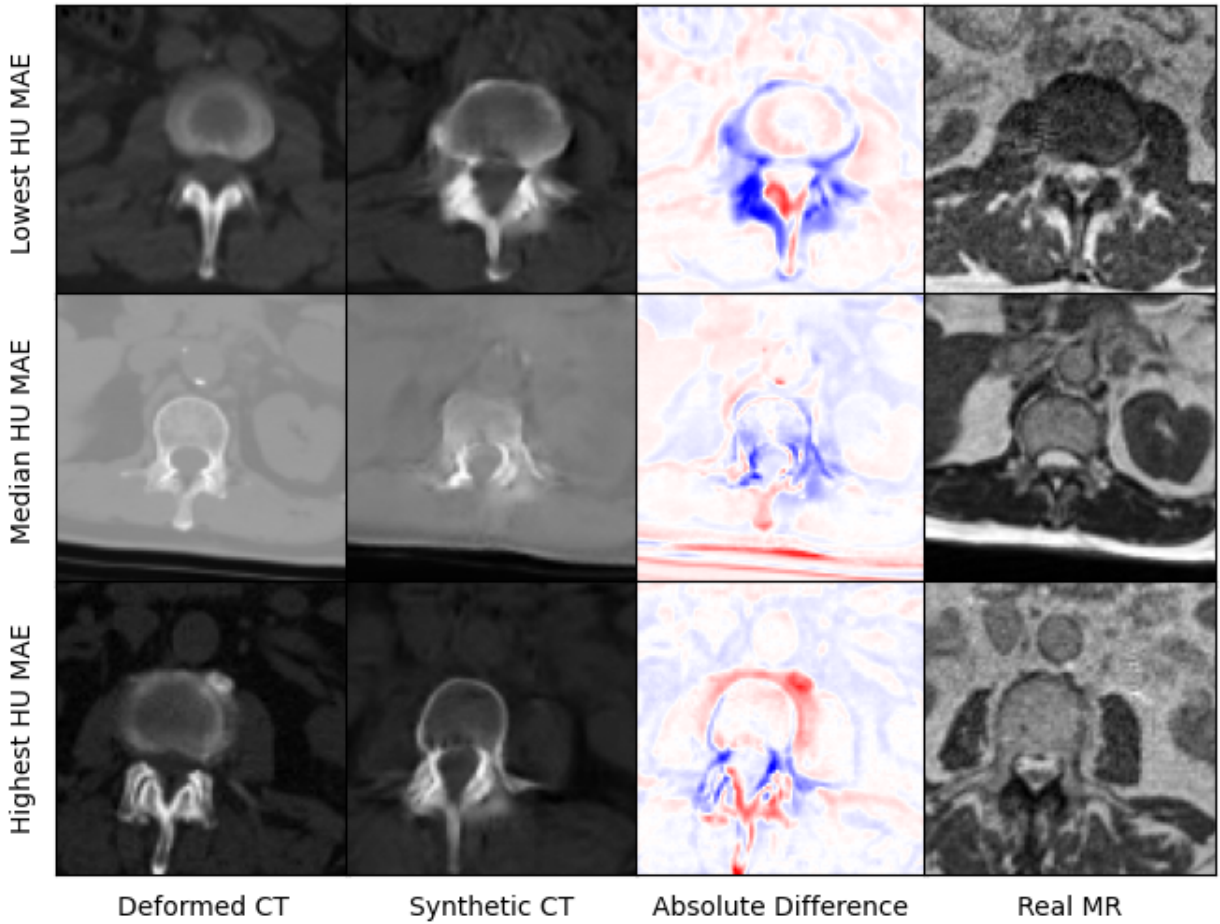

**Figure 5:** Zoomed in panel of spinal region of interest. We compare sCT spinal reconstruction on lowest, median, and highest HU MAE patients.

Some limitations of our approach include limited frame-to-frame cohesion on the axial plane. We observed that although primary image structures and features remained fixed, there were some axial frame-to-frame shifts which occurred along the extremities of each scan. These shifts may have occurred because the CycleGAN architecture trains on a single image at a time as opposed to a whole volume. Therefore, the model has more difficulty learning frame to frame continuance. Another observed limitation to our method is that our model has difficulty predicting HU values for high-intensity regions. For example, in **Figure 5**, we see that for all 3 image examples our model had difficulty reconstructing and predicting spine intensity values. We believe that this is due to the intrinsic distribution of Houndsfield Unit values. Bone HU values are typically above 700, whereas all other tissue HU values fall between -100 and 300 HU. Thus, distribution imbalances may have led to difficulties predicting HU values for high-intensity regions. Finally, we believe that some distribution-matching losses may have caused anomalously shaped structures in certain already hard-to-predict regions such as the spine. For example, the spinal structure of all three sCTs in **Figure 5** differ from their dCT ground truths. Given their relative similarity to each other, we believe that these differences may correspond to hallucinated features caused by distribution-matching losses, a phenomenon previously observed in Cycle-GAN based image translation.[30]

Future work will involve two key advancements. Firstly, larger and higher quality datasets must be created. Our study shows that despite similar architectures, dataset scale improved our results in comparison to prior studies. Additionally, we believe that on top of scaling dataset, some incorporation of newer DL architectures could improve results. For example, although some work has already been done on MRI to CT conversion using diffusion models,[31] given the remarkable performance of diffusion models in other image-processing domains,[32] these models are promising for a synthetic CT generation task.

## Conclusions

Accuracy of deep learning based synthetic CT generation using setup scans on MR-Linacs was adequate for dose calculation/optimization. This can enable MR-only treatment planning workflows on MR-Linacs, thereby increasing the efficiency of simulation and adaptive planning for MRgART.

## Acknowledgements

This work was supported in part by funding from the Dartmouth-Hitchcock Medical Center Department of Medicine through the Scholarship Enrichment in Academic Medicine program.

## Ethical approval

This study was conducted retrospectively using anonymized patient data. Ethical approval for this retrospective analysis was obtained from the Dartmouth-Hitchcock Institutional Review Board.